\documentclass{appolb}
\usepackage{graphicx}
\begin{document}
\title{Spectral Functions and Pseudogap in Models 
of Strongly Correlated Electrons%
\thanks{Presented at the Strongly Correlated Electron Systems 
Conference, Krak\'ow 2002}%
}
% Authors and Affiliations

\author{P. Prelov\v sek and A. Ram\v sak
\address{Faculty of Mathematics and
Physics, University of Ljubljana,\\ 1111 Ljubljana, Slovenia \\
J. Stefan Institute, University of Ljubljana, 1111 Ljubljana,
Slovenia}
}
\maketitle
   
% Abstract

\begin{abstract}
The theoretical investigation of spectral functions and pseudogap in
systems with strongly correlated electrons is discussed, with the
emphasis on the single-band $t$-$J$ model as relevant for
superconducting cuprates. The evidence for the pseudogap features from
numerical studies of the model is presented. One of the promising
methods to study spectral functions is the method of equations of
motion. The latter can deal systematically with the local constraints
and projected fermion operators inherent for strongly correlated
electrons. In the evaluation of the self energy the decoupling of spin
and single-particle fluctuations is performed. In an undoped
antiferromagnet the method reproduces the selfconsistent Born
approximation (SCBA). For finite doping the approximation evolves into a
paramagnon contribution which retains large incoherent contribution in
the hole part. On the other hand, the contribution of longer-range
spin fluctuations is essential for the emergence of the pseudogap. The
latter shows up at low doping in the effective truncation of the large
Fermi surface, reduced electron density of states and at the same time
reduced quasiparticle density of states at the Fermi level.
\end{abstract}

    
\section{Introduction}

One of the central questions in the theory of strongly correlated
electrons is the nature of the ground state and of low energy
excitations. Experiments in many novel materials with correlated
electrons \cite{imad} reveal even in the 'normal' metallic state
striking deviations from the usual Fermi-liquid universality as given
by the phenomenological Landau theory involving quasiparticles (QP) as
the well defined fermionic excitations even in the presence of Coulomb
interactions. The focus has been and still remains on superconducting
cuprates where there exists now an abundant and consistent experimental
evidence for very anomalous low-energy properties \cite{imad}, besides
the most evident open question of the origin of the high-$T_c$
superconductivity.

In particular, the attention in the last decade has been increasingly
devoted to the underdoped cuprates, where experiments reveal
characteristic 'pseudogap' temperatures, which show up crossovers
where particular properties change quantitatively.  As schematically
presented in Fig.~1, there seems to be an indication for two crossover
scales $T^*$ \cite{batl} and $T_{sg}$ \cite{imad}. The existence of both is
still widely debated, in particular whether both could be a
manifestation of the same underlying mechanism. Nevertheless we refer
to them (as usually in the literature) as the (larger) pseudogap scale
$T^*$ and the spin-gap scale $T_{sg}$ for the lower one.
\begin{figure}[!ht]
\begin{center}
\includegraphics[width=0.45\textwidth]{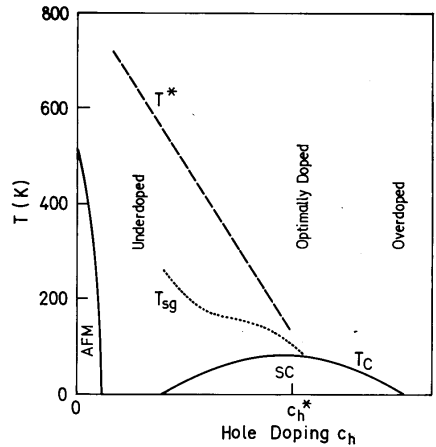}
\end{center}
\caption{Schematic electronic phase diagram of cuprates}
\label{fig1}
\end{figure}

The $T^*$ scale \cite{batl} shows up clearly as the maximum of the
spin susceptibility $\chi_0(T=T^*)$ \cite{torr}.  The in-plane
resistivity $\rho(T)$ is linear $\rho \propto T$ for $T>T^*$ and
decreases more steeply for $T<T^*$. The same $T^*$ appears in the
anomalous $T$-dependence of the Hall constant $R_H(T)$ for $T<T^*$
\cite{imad}. For the theoretical considerations the most
straightforward signature of pseudogap is the reduction of the
specific heat coefficient $\gamma=C_V/T$ below $T\sim T^*$ in the
underdoped cuprates \cite{lora}. Namely, within normal Fermi liquids
$\gamma$ directly measures the quasiparticle (QP) density of states.
The spin-gap crossover $T_{sg}$ has been identified in connection with
the decrease of the NMR relaxation rate $1/T_1$ for $T<T_{sg}$
\cite{imad}, related to the reduction of low-energy spin excitations.
Even more striking is the observation of the leading-edge shift
\cite{camp} in the angle-resolved photoemission spectroscopy (ARPES)
measurements at $T>T_c$, a feature interpreted as a $d$-wave SC gap
persisting within the normal phase.

It seems to some extent plausible that the $T^*$ crossover is related
to the onset of short-range antiferromagnetic (AFM) correlations for
$T<T^*$, since $\chi(T)$ in an undoped AFM has a maximum at $T^*\sim
2J/3$ the temperature corresponding to a gradual transition from a
disordered paramagnet to the one with short-range AFM correlations. In
this contribution we concentrate on our theoretical results which
confirm and explain the existence of the pseudogap in the model
relevant for cuprates, i.e. the $t$-$J$ model on a 2D square
lattice. It should be however pointed out that there there are also
several alternative theoretical proposals, which do not directly
invoke AFM spin correlations.

The single-particle spectral function $A({\bf k},\omega)$ and its
properties are of crucial importance, since their full knowledge would
essentially clarify most open questions concerning the anomalous
character of strongly correlated electron systems.  In recent years
there has been an impressive progress in ARPES experiments
\cite{camp,fuji,imad} (in particular for cuprate materials) which in
principle yield a direct information on $A({\bf k},\omega)$. In most
investigated Bi$_2$Sr$_2$CaCu$_2$O$_{2+\delta}$ (BSCCO) \cite{camp}
ARPES shows quite a well defined large Fermi surface (FS) in the
overdoped and optimally doped samples at $T>T_c$, whereby the
low-energy behavior with increasing doping in the overdoped regime
qualitatively approaches (but does not in fact reach) that of the
normal Fermi-liquid with underdamped quasiparticle (QP)
excitations. On the other hand, in the underdoped BSCCO QP dispersing
through the Fermi surface (FS) are resolved by ARPES only in parts of
the large FS, in particular along the nodal $(0,0)$-$(\pi,\pi)$
direction \cite{mars,camp}, indicating that the rest of the large FS
is truncated \cite{norm}, i.e. either fully or effectively gaped.  At
the same time near the $(\pi,0)$ momentum ARPES reveals a hump at
$\sim 100$~meV \cite{mars,camp}, which is consistent with large
pseudogap scale $T^*$. Spectral properties for
La$_{2-x}$Sr$_x$Cu$_2$O$_4$ (LSCO), as revealed by ARPES \cite{fuji}
appear to some extent different from BSCCO, presumably due to the
crucial role of stripe structures in the LSCO in the regime of
intermediate doping. Still they again reveal a truncated FS at
low-doping and even the existence of QP along the nodal direction.

The prototype single-band model relevant for cuprates which takes
explicitly into account strong correlations is the $t$-$J$ model,
derived originally by Chao, Spa\l{}ek and Ole\' s \cite{chao}
\begin{equation}
H=-\sum_{i,j,s}t_{ij} \tilde{c}^\dagger_{js}\tilde{c}_{is}
+J\sum_{\langle ij\rangle}({\bf S}_i\cdot {\bf S}_j-\frac{1}{4}
n_in_j) , \label{eq1}
\end{equation}
where fermionic operators are projected ones not allowing for the
double occupancy of sites, i.e., 
\begin{equation}
\tilde{c}^\dagger_{is}= (1-n_{i,-s}) c^\dagger_{is}. \label{eq2}
\end{equation}
Longer range hopping appears to be important for the proper
description of spectral function in cuprates, in particular it is
invoked to explain the difference between electron-doped and
hole-doped materials both in the shape of the FS at optimum doping
materials \cite{tohy} as well as for the explanation of the ARPES of
undoped insulators \cite{well,tohy}, we consider besides $t_{ij}=t$
for the n.n. hopping also $t_{ij}=t'$ for the n.n.n. hopping on a
square lattice. Note that $t'<0$ for the hole-doped cuprates.

There have been so far numerous theoretical studies of the $t$-$J$
model on square lattice and related Hubbard model at large Coulomb
repulsion $U \gg t$ , as relevant to cuprates, using both analytical
approaches as well as numerical techniques for finite size
systems. Still analytical approximations to spectral properties have
proved to be very delicate, in particular with respect to the question
of the emerging pseudogap at lower doping. So there are much fewer
studies which give some answers on latter questions within
microscopic models close to cuprates. The importance of AFM spin
correlations for the emergence of the (large) pseudogap is found
in the numerical studies \cite{preu,jpspec,jprev} and in
phenomenological model studies \cite{chub}. The renormalization
group studies of the Hubbard model \cite{zanc} also reveal the
instability of the normal Fermi liquid close to the half-filled band
(insulator) and a possible truncation of the Fermi surface.

In the following we describe some evidence for the pseudogap within
the $t$-$J$ model obtained via finite-size studies and a novel
approach to spectral functions using the method of equations of
motion.

\section{Evidence for the pseudogap from numerical studies}

We introduced some years ago a numerical method, i.e. the
finite-temperature Lanczos method (FTLM) \cite{jplanc,jprev}, which is
particularly useful for studying finite-size model systems of
correlated electrons at $T>0$. The technical advantage of the method
is that it is comparable in efficiency to ground state
calculations. $T>0$ results for static and dynamical quantities are of
interest in themselves, allowing to follow the $T$-variation of
properties whereby some of them are meaningful only at $T>0$,
e.g. the entropy, the specific heat, the d.c.resistivity etc. On the other
hand, the usage of finite but small $T>0$ represents a proper approach
to more reliable ground state calculations in small systems.

\begin{figure}[!ht]
\begin{center}
\includegraphics[angle=-90,width=0.7\textwidth]{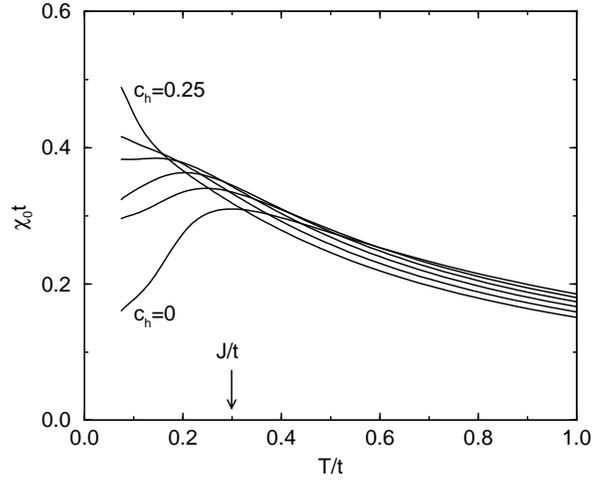}
\end{center}
\caption{Uniform susceptibility $\chi_0(T)$ at several hole doping 
$c_h$. Results are for $J/t=0.3$. }
\label{fig2}
\end{figure}
The most straightforward evidence for a pseudogap within the planar
$t$-$J$ model allowing the comparison with experiments appears in the
uniform static spin-suceptibility $\chi_0(T)$
\cite{jpterm,jprev}. Results for various hole concentrations
$c_h=N_h/N$ in a system with $N=20$ sites are presented in Fig.~2. It
is evident that the maximum $T^*$ being related to the AFM exchange
$T^* \sim 2J/3$ in an undoped AFM gradually shifts down with doping
and finally disappears at 'critical' $c_h =c_h^* \sim 0.15$. Obtained
results are qualitatively as well as quantitatively consistent with
experiments in cuprates, e.g. LSCO \cite{torr}.
\begin{figure}[!ht]
\begin{center}
\includegraphics[width=0.7\textwidth]{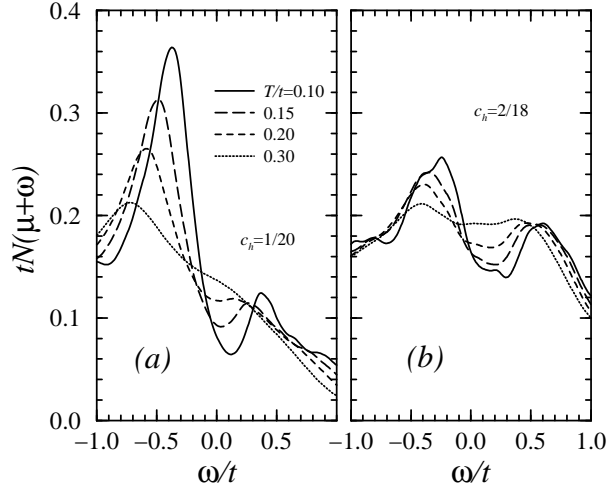}
\end{center}
\caption{The DOS ${\cal N}(\mu+\omega)$ at various $T\leq J$ for hole
doping: a) $c_h=1/20$ and b) $c_h=2/18$. }
\label{fig3}
\end{figure}

Another quantity relevant for comparison with the analytical approach
further on is the single-particle density of states (DOS) ${\cal
N}(\omega)$ \cite{jpspec,jprev}. We present in Fig.~3 numerical
results for DOS \cite{prel2} as a function of $T$ for systems with
$N=18, 20$ sites and two lowest nonzero meaningful hole concentrations
$c_h \sim 0.05, 0.11$. At smallest $c_h=0.05$ there is a pronounced
pseudogap at $\omega \sim 0$ which closes with increasing $T \sim T^*
\sim J$. This again indicates the relation of this pseudogap with the
AFM short-range correlations which dissolve for $T>J$.  On the other
hand, the pseudogap closes also on increasing doping since it becomes
barely visible at $c_h \sim 0.11$.

\section{Spectral functions: Equation-of-motion approach}

In our analytical approach we analyze the electron Green's function
directly for projected fermionic operators \cite{prel,prel1},
\begin{equation}
G({\bf k},\omega)= \langle\!\langle \tilde c_{{\bf k} s}; \tilde
c^{\dagger}_{{\bf k} s} \rangle\!\rangle _{\omega} = -i \int_0^{\infty}
{\rm e}^{i(\omega+\mu ) t}\langle \{ \tilde c_{{\bf k} s}(t) ,
\tilde c^{\dagger}_{{\bf k} s} \}_+  \rangle {\rm d}t, 
\label{eq3}
\end{equation}
which is equivalent to the usual propagator within the allowed basis
states of the model. In the EQM method \cite{zuba} one uses relations
for general correlation functions
\begin{equation}
\omega \langle \!\langle A;B \rangle \!\rangle_\omega = \langle
 \{A,B\}_+\rangle + \langle \!\langle [A,H]; B \rangle
\!\rangle_\omega 
\label{eq4}
\end{equation}
applying them to the propagator $G({\bf k},\omega)$ \cite{prel,prel1}
one can represent the latter in the form
\begin{equation}
G({\bf k},\omega)= \frac{\alpha}{\omega+\mu -\zeta_{\bf k} - \Sigma({\bf
k},\omega) }, \label{eq5}
\end{equation}
where $\alpha$,$\zeta_{\bf k}$ can be expressed in terms of
commutators. It is important to notice that the renormalization $\alpha<1$
is already the consequence of the projected basis,
\begin{equation}
\alpha = \frac{1}{N}\sum_i \langle \{\tilde c_{i s},\tilde
 c^{\dagger}_{i s}\}_+ \rangle = \frac {1}{2}
 (1+c_h), \label{eq6}
\end{equation}
while $\zeta_{\bf k}$ represents the 'free' propagation emerging from
EQM,
\begin{equation}
  \zeta_{\bf k}= \frac{1}{\alpha} \langle \{[\tilde c_{{\bf k} s},H],
  \tilde c^{\dagger}_{{\bf k} s}\}_+\rangle -\bar \zeta
= -4 \eta_1 t \gamma_{\bf k}
  -4 \eta_2 t' \gamma'_{\bf k}, \label{eq7}
\end{equation}
where  $\eta_j = \alpha + \langle {\bf S}_0 \!\cdot\! {\bf S}_j
\rangle/\alpha$ and $\gamma_{\bf k}=(\cos
k_x+\cos k_y)/2$, $\gamma_{\bf k}'=\cos k_x \cos k_y$.

The central quantity for further consideration is the self energy
\begin{equation}
\Sigma({\bf k},\omega) = \langle\!\langle C_{{\bf k}s}; C^+_{{\bf k}s}
\rangle\!\rangle_\omega^{\rm irr} /
\alpha, \qquad iC_{{\bf k}s}=[\tilde c_{{\bf k} s},H]-\zeta_{\bf k}
\tilde c_{{\bf k} s}, \label{eq8}
\end{equation}
and only the 'irreducible' part of the correlation function should be
taken into account in the evaluation of $\Sigma$.  EQM enter in the
evaluation of $\zeta_{\bf k}$ but even more important in $C_{{\bf
k}s}$. We express the commutator in variables appropriate for a
paramagnetic metallic state with $\langle {\bf S}_i \rangle =0$,
and we get 
\begin{eqnarray}
[\tilde c_{{\bf k} s},H]&=& [(1-\frac {c_e}{2}) \epsilon^0_{\bf k} - J
c_e]\tilde c_{{\bf k} s} + \nonumber \\
& &+ \frac{1}{\sqrt{N}} \sum_{\bf q} m_{\bf k q}
\bigl[ s S^z_{\bf q} \tilde c_{{\bf k}-{\bf q},s} + S^{\mp}_{\bf q}
\tilde c_{{\bf k}-{\bf q},-s} - \frac{1}{2} \tilde n_{\bf q} \tilde
c_{{\bf k}-{\bf q}, s}\bigr],
\label{eq9}
\end{eqnarray}
where $\epsilon^0_{\bf k}=-4t\gamma_{\bf k}-4t'\gamma'_{\bf k}$ is the
bare band energy, $\tilde n_i=n_i-c_e$ and $m_{\bf k q}$ is an effective
spin-fermion coupling,
\begin{equation}
 m_{\bf k q}=2J \gamma_{\bf q} + \epsilon^0_{{\bf k}-{\bf q}}.
\label{eq10}
\end{equation}
One important achievement of the EQM approach is that it naturally
leads to an effective coupling (not directly evident within the
$t$-$J$ model) between fermionic and spin degrees of freedom, which
are essential for proper description of low-energy physics in
cuprates. Such a coupling is e.g. assumed in phenomenological models
as the spin-fermion model \cite{mont,chub}. The essential difference
in our case is that $m_{\bf k q}$ is strongly dependent on ${\bf k}$
and ${\bf q}$ just in the vicinity of most relevant 'hot' spots.

\section{Self energy}

\subsection{Undoped AFM and short range AFM fluctuations}

It is quite helpful observation that in the case of an undoped AFM our
treatment of $\Sigma$ and the spectral function reproduces quite
successful SCBA equations \cite{kane} for the Green's function of a
hole in an AFM. If we write EQM in the coordinate space for $\tilde
c_{is}$ assuming the N\' eel state as the reference $n_{is}=\pm 1$, we
get by considering only the $t$ term,
\begin{equation}
i\frac{d}{dt} \tilde c_{is} \sim -t \sum_{j~ n.n.i} (S^\mp_i+ S^\mp_j)
\tilde c_{j,-s}, \label{eq11}
\end{equation}
where we have also formally replaced $\tilde c_{js}=\tilde c_{j,-s}
S^\mp_j $. We note the similarity of Eq.~(\ref{eq11}) to the effective
spin-holon coupling within the SCBA approach. So we can follow the
procedure of the evaluation of $\Sigma_{\rm AFM}({\bf k}, \omega)$
within the SCBA in the linearized magnon theory \cite{kane} 
\begin{equation}
\Sigma_{\rm AFM}({\bf k},\omega)= \frac{1}{N} \sum_{\bf q}
M_{\bf kq}^2 G({\bf k}-{\bf q},\omega+\omega_{\bf q}), 
\label{eq12}
\end{equation}
where $\omega_{\bf q}$ is the magnon dispersion and 
$M_{\bf kq} = 4t (u_{\bf q} \gamma_{{\bf k}-{\bf q}}+v_{\bf q}
\gamma_{\bf k})$ is the holon-magnon coupling which in general strong 
$ \propto t$, but vanishes near the AFM wavevector ${\bf q}={\bf
Q}=(\pi,\pi)$.

The advantage of the representation of EQM, Eq.~(\ref{eq11}),
explicitly in spin and fermionic variables is that it allows the
generalization to finite doping $c_h>0$. We assume that spin
fluctuations remain dominant at the AFM wavevector ${\bf Q}$ with the
characteristic inverse AFM correlation length $\kappa=1/\xi_{AFM}$.
The latter seems to be the case for BSCCO as well as
YB$_2$Cu$_3$O$_{6+x}$, but not for LSCO with pronounced stripe and
spin-density structures \cite{imad}. For BSCCO and YBCO it is sensible
to divide the spin fluctuations into two regimes with respect to
$\tilde {\bf q}= {\bf q}-{\bf Q}$: a) For $\tilde q>\kappa$ spin
fluctuations are paramagnons, they are propagating like magnons and
are transverse to the local AFM short-range spin.  b) For
$\tilde q<\kappa$ spin fluctuations are essentially not propagating
modes but critically overdamped so deviations from the long range
order are essential.

At finite doping case we therefore generalize (at $T=0$)
Eq.~(\ref{eq19}) into the paramagnon contribution,
\begin{equation}
\Sigma_{\rm pm}({\bf k},\omega)= \!\frac{1}{N} \!\!\!\sum_{q,\tilde q>
\kappa}  \bigl[ M_{\bf kq}^2  
G^-({\bf k}-{\bf q},\omega+\omega_{\bf q}) +  M_{{\bf
k}+{\bf q},{\bf q}}^2 G^+({\bf k}+{\bf q},\omega-\omega_{\bf q})],
\label{eq22}
\end{equation}
where $G^\pm({\bf k},\omega)$ refer to the electron ($\omega>0$) and
hole-part ($\omega<0$) of the propagator, respectively.  We are
dealing in Eq.~(\ref{eq22}) with a strong coupling theory due to $t >
\omega_{\bf q}$ and a selfconsistent calculation of $\Sigma_{\rm pm}$
is required \cite{prel1}.  Also, resulting $\Sigma_{\rm pm}({\bf k},\omega)$ and
$A({\bf k},\omega)$ as are at low doping quite asymmetric with respect
to $\omega = 0$, since $G^-\propto (1-c_h)/2 \sim 1/2$ while $G^+
\propto c_h$.

\subsection{Longitudinal spin fluctuations}

At $c_h>0$ the electronic system is in a paramagnetic state without an
AFM long-range order and besides the paramagnon excitations also the
coupling to longitudinal spin fluctuations become crucial. The latter
restore the spin rotation symmetry in a paramagnet and EQM
(\ref{eq9}) introduce such a spin-symmetric coupling.  Within a
simplest approximation that the dynamics of fermions and spins is
independent, we get
\begin{equation}
\Sigma_{\rm lf}({\bf k},\omega) =\frac{1}{\alpha} \sum_{\bf q} 
\tilde m^2_{\bf k q}
\int \!\! \int \frac{{\rm d}\omega_1 {\rm d}\omega_2}{\pi} g(\omega_1,\omega_2)
\frac{\tilde A({{\bf k}-{\bf q}},\omega_1) \chi''({\bf q},\omega_2)}
{\omega-\omega_1-\omega_2 },\label{eq14}
\end{equation}
where $ g(\omega_1,\omega_2)= (1/2)[{\rm th}(\beta\omega_1/2)+{\rm
cth}(\beta\omega_2/2)]$ and $\chi$ is the dynamical spin
susceptibility.  Quite analogous treatment has been employed
previously in the Hubbard model \cite{kamp} and more recently within
the spin-fermion model \cite{chub,schm}.

If we want to use the analogy with the spin-fermion Hamiltonian the
effective coupling parameter $\tilde m_{\bf kq}$ should satisfy
$\tilde m_{{\bf k},{\bf q}} = \tilde m_{{\bf k}-{\bf q},-{\bf q}}$
which is in general not the case with the form Eq.~(\ref{eq10}),
therefore we use further on instead the symmetrized coupling
\begin{equation}
\tilde m_{\bf kq}= 2J \gamma_{\bf q}+ \frac{1}{2} 
(\epsilon^0_{{\bf k}-{\bf q}}+\epsilon^0_{\bf k}). \label{eq15}
\end{equation}
In contrast to previous related studies of spin-fermion coupling
\cite{kamp,chub,schm}, however, $\tilde m_{\bf kq}$ is strongly
dependent on both ${\bf q}$ and ${\bf k}$. It is essential that in the
most sensitive parts of the FS, i.e., along the AFM zone boundary
('hot' line) where $k=|{\bf Q}-{\bf k}|$, the coupling is in fact
quite modest determined solely by $J$ and $t'$. Also, in the regime
close to that of quasistatic $\chi({\bf q},\omega)$ the simplest and
also quite satisfactory approximation is to insert for $\tilde A$ the
unrenormalized $A^0$ \cite{chub}, the latter corresponding in our case
to the spectral function without $\Sigma_{\rm lf}$ but with
$\Sigma=\Sigma_{\rm pm}$.

In the present theory spin susceptibility $\chi({\bf q},\omega)$ is
taken as an input.  The system is close to the AFM instability, so we
assume spin fluctuations of the overdamped form
\cite{mont}
\begin{equation}
\chi''({\bf q},\omega) \propto\frac{ \omega}
{(\tilde q^2 +\kappa^2) (\omega^2+\omega_\kappa^2)}. \label{eq16}
\end{equation}
Nevertheless, the appearance of the pseudogap and the form of the FS are not
strongly sensitive to the particular form of $\chi''({\bf q},\omega)$
at given characteristic $\kappa$ and $\omega_\kappa$.

\section{Pseudogap}

We first establish some characteristic features of the
pseudogap and the development of the FS following a simplified
analysis. We note that $\Sigma_{\rm pm}$ 
induces a large incoherent component in the spectral
functions at $\omega \ll 0$ and renormalizes the 
effective QP band relevant to the
behavior at $\omega \sim 0$ and at the FS \cite{prel1}.
It can also lead to a transition of a large FS into a small
hole-pocket-like  FS at very small $c_h<c_h^* \ll 1$.
Nevertheless, the pseudogap can appear only via $\Sigma_{\rm lf}$.
Therefore we here take into account $\Sigma_{\rm pm}$ 
only via an effective band $\epsilon_{\bf k}^{\rm ef}$.
The input spectral function  for $\Sigma_{\rm lf}$ is thus
\begin{equation}
A^0_{\rm ef}({\bf k},\omega)=\alpha Z^{\rm ef}_{\bf k} \delta(\omega +\mu
-\epsilon_{\bf k}^{\rm ef}). \label{eq17}
\end{equation}
We restrict our discussion to $T=0$ and to the regime of intermediate
(not too small) doping, where $\epsilon_{\bf k}^{\rm ef}$ defines a
large FS. The simplest case is the quasi-static and single-mode
approximation (QSA) which is meaningful if $\omega_\kappa \ll t,
\kappa \ll 1$, where we get
\begin{equation} 
G^{QSA}({\bf k},\omega)= \frac{\alpha Z^{\rm ef}_{\bf k} (\omega -
\epsilon^{\rm ef}_{{\bf k}-{\bf Q}} )} {(\omega - \epsilon^{\rm ef}_{{\bf
k}-{\bf Q}})(\omega - \epsilon^{\rm ef}_{\bf k}) - \Delta^2_{\bf k} }.
\label{eq18}
\end{equation} 
The spectral function shows in this approximation two branches of
$E^\pm$, separated by the gap which opens along the AFM zone boundary
${\bf k}={\bf k}_{AFM}$ and the relevant (pseudo)gap scale is
\begin{equation}
\Delta^{PG}_{\bf k} = |\Delta_{{\bf k}_{AFM}}| = 
\frac{Z^{\rm ef}_{\bf k}}{2} 
\sqrt{ r_s} |2J - 4 t' {\rm cos}^2 k_x|. \label{eq19}
\end{equation}
$\Delta^{PG}_{\bf k}$ does not depend on $t$, but rather on smaller
$J$ and in particular $t'$. For $t'<0$ the gap is largest at
$(\pi,0)$, consistent with experiments \cite{mars,camp}. 

QSA yileds a full gap corresponding, e.g., to the case of a long-range
AFM state. Within the simplified effective band approach,
Eq.~(\ref{eq17}), it is not difficult to evaluate numerically
$\Sigma_{\rm lf}$ beyond the QSA, by taking explicitly $\chi''({\bf
q},\omega)$, Eq.~(\ref{eq16}), with $\kappa>0$ and $\omega_\kappa \sim
2J\kappa$. For illustration, we present results characteristic for the
development of spectral functions varying two most sensitive
parameters $\kappa$ and $\mu$, which both simulate the variation with
doping, e.g. one can $\kappa$ take in accordance with experiments
\cite{imad} and numerical results on the $t$-$J$ model \cite{dago} as
$\kappa \sim \sqrt{c_h}$.

In Fig.~3 we present results for $A({\bf k},\omega=0)$ at $T=0$ for a
broad range of $\kappa=0.01 - 0.6$. Curves in fact display the
effective FS determined by the condition $G^{-1}({\bf k}_F,0)\sim
0$. At the same time, intensities $A({\bf k},\omega=0)$ correspond to
the QP weight $Z_F$ at the FS. At very small $\kappa=0.01$ we see a
small (hole-pocket) FS which follows from the QSA,
Eq.~(\ref{eq18}). Already $\kappa \sim 0.05$ destroys the 'shadow' side
of the pocket, i.e., the solution $G^{-1}=0$ on the latter side
disappears. On the other hand, in the gap emerge now QP solutions with
very weak $Z_F \ll 1$ which reconnect the FS into a large one. We are
dealing nevertheless with effectively truncated FS with well developed
arcs. The effect of larger $\kappa$ is essentially to increase $Z_F$
in the gapped region, in particular near $(\pi,0)$. Finally, for large
$\kappa = 0.6$ corresponding in cuprates to the optimal doping or
overdoping, $Z_F$ is essentially only weakly decreasing towards
$(\pi,0)$ and the FS is well pronounced and concave as naturally
expected for $t'<0$.

\begin{figure}[!ht]
\begin{center}
\includegraphics[width=0.6\textwidth]{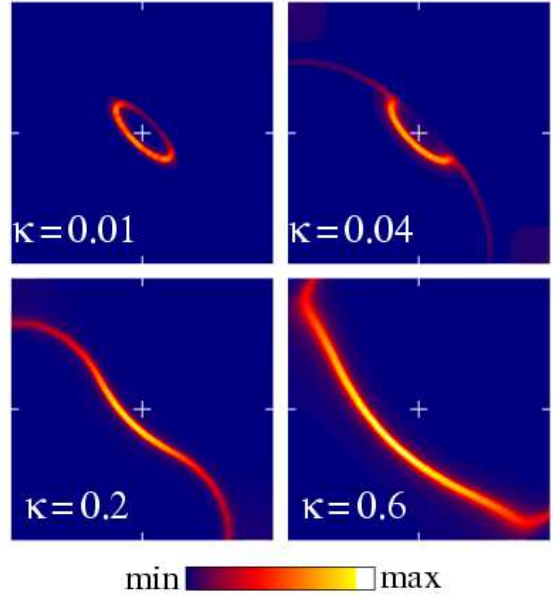}
\end{center}
\caption{Contour plot of spectral functions $A({\bf k},\omega=0)$ at
$T=0$ for various $\kappa$ in one quarter of the Brillouin zone.}
\label{fig4}
\end{figure}

We note in Fig.~4 that except at extreme $\kappa = 0.01$ we get a
large FS, whereby in the 'underdoped' regime $\kappa < \kappa^* \sim
0.5$ the QP weight $Z_F$ is substantial only within pronounced 'arcs'
and very small along the 'gapped' FS where $Z_F \ll 1$. Still, such a
situation corresponds to a Fermi liquid, although a very strange one,
where QP excitations exist everywhere along the FS and hence determine
the low-energy properties of the 'normal' metallic state.

It is quite remarkable to notice that in spite of $Z_F \ll 1$ the QP
velocity $v_F$ is not diminished within the pseudogap. In fact it can
be even enhanced, as seen in in Fig.~5 where the contour plot of
$A({\bf k},\omega)$ is shown. Again, it is well evident in Fig.~5 that
QP is well defined at the FS, while it becomes fuzzy at $\omega \neq
0$ merging with the solutions $E_{\bf k}^\pm$, respectively, away from
the FS. The effect of large $v_F$ in the pseudogap, which is essential
for the low-$T$ thermodynamics, can be only explained with a crucial
${\bf k}$ dependence of $\Sigma$. Assuming that ${\bf k}$ enters only
via $\epsilon({\bf k})$ we can express $v_F$ renormalization as
\begin{equation}
\frac{v_F}{v^{\rm ef}_{\bf k}}= (1+ \frac{\partial\Sigma'} {\partial
\epsilon} )\frac{Z_F}{Z^{\rm ef}}, \qquad
\frac{\partial\Sigma'}{\partial \epsilon}\Bigr|_{\omega=0} \sim
\frac{\Delta^2}{w(\omega_\kappa+w)}
. \label{eq20}
\end{equation}
While $Z_F/Z^{ef}\ll 1$ in the same case, $\partial\Sigma'/\partial
\epsilon$ compensates or even leads to an enhancement of $v_F$.  In
the case $\omega_\kappa w \ll \Delta^2$ we get
\begin{equation}
\frac{v_F}{v^{\rm ef}_{\bf k}} \sim \frac{\omega_\kappa}{w} \sim
\frac{2J}{v_{\bf k}}. \label{eq21}
\end{equation}
Final $v_F$ is therefore not strongly renormalized, since $2J$ and
$v^{\rm ef}_{\bf k}$ are of similar order. Furthermore, $v_F$ is
enhanced in the parts of FS where $v^{\rm ef}_{\bf k}$ is small, in
particular near $(\pi,0)$ point. The situation is thus very different
from 'local' theories where $\Sigma({\bf k},\omega) \sim
\Sigma(\omega)$ and the QP velocity is governed only by $Z_F$. In our
case the 'nonlocal' character of $\Sigma({\bf k},\omega)$ is essential
in order to properly describe QP within the pseudogap region.

\begin{figure}[!ht]
\begin{center}
\includegraphics[width=0.5\textwidth]{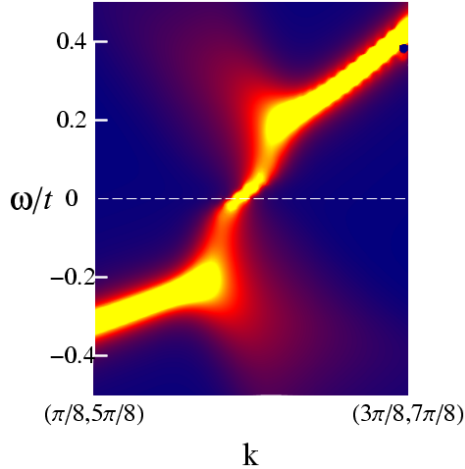}
\end{center}
\caption{Contour plot of spectral functions $A({\bf k},\omega=0)$ 
across the FS in the pseudogap regime.}
\label{fig5}
\end{figure}

>From $A({\bf k},\omega)$ we can calculate the DOS ${\cal N}
(\omega)=(2/N) \sum_{\bf k} A({\bf k},\omega)$. At the Fermi energy
$\omega \sim 0$ and low $T$ one can express the DOS also as (assuming
the existence of QP around the FS),
\begin{equation}   
{\cal N}(0) \sim \frac{\alpha} {2\pi^2} \oint \frac{{\rm d}S_F Z({\bf k}_F)
}{v({\bf k}_F)}. \label{22}
\end{equation}   
The contribution will come mostly from FS arcs near the zone diagonal
while the gapped regions near $(\pi,0)$ will contribute less due to
$Z({\bf k}_F) \ll 1$.  Results in Fig.~6(a) show the development of
${\cal N}(0)$ with $\kappa$. As expected the DOS decreases with
decreasing $\kappa$ simulating the approach to an undoped AFM.  The
DOS is measured in cuprates via angle integrated photoemission
spectroscopy, e.g. for LSCO in \cite{ino}, as well as via the scanning
tunneling microscopy (STM) \cite{renn}. Our results are qualitatively
consistent with these experiments (showing ${\cal N}(0)$ scaling with
doping) as well as with numerical results on the $t$-$J$ model
\cite{jpspec,jprev}, as shown in Fig.~3.  It is well possible that
within photoemission experiments the matrix elements are essential, so
we present in Fig.~6(a) also results weighted by the matrix element
$w({\bf k})= (\cos k_x - \cos k_y)^2$, which originates from the
interplanar hopping as proposed for the $c$-axis conductivity
\cite{ioff}. Such weighted results show a stronger dependence on
doping due to enhanced influence of the pseudogap region near
$(\pi,0)$, which could be even closer to experimental findings
\cite{ino}.
In Fig.~6(b) we show also the average $Z_{\rm av}$ along the FS, as
well as the QP DOS, defined as
\begin{equation}   
{\cal N}_{QP}= \frac{1}{2\pi^2} \oint 
\frac{{\rm d}S_F}{v({\bf k})}. \label{eq23}
\end{equation}   

The decrease $Z_{\rm av}$ has similar dependence at the DOS ${\cal
N}(0)$.  It is however quite important to notice that smaller doping
(decreasing $\kappa$) leads also to a decrease of ${\cal N}_{QP}$.
This is consistent with the observation of the pseudogap also in the
specific heat in cuprates \cite{lora}, since ${\cal N}_{QP} \propto
\gamma=C_V/T$. We note here that such a behavior is not at all evident
in the vicinity of a metal-insulator transition \cite{imad}. Namely,
in a Fermi liquid with (nearly constant) large FS one can drive the
metal-insulator transition by $Z_{\rm av} \to 0$. Within an assumption
of a local character $\Sigma(\omega)$ this would also lead to $v_F \to
0$ and consequently via Eq.~(\ref{eq23}) to ${\cal N}_{QP} \to
\infty$. Clearly, the essential difference in our case is that within
the pseudogap regime $\Sigma({\bf k},\omega)$ is nonlocal, allowing
for a large (not reduced or even enhanced) velocity within the
pseudogap regime, hence a simultaneous decrease of ${\cal N}(0)$ and
${\cal N}_{QP}$.

\begin{figure}[htb]
\begin{center}
\includegraphics[width=0.5\textwidth,angle=-90]{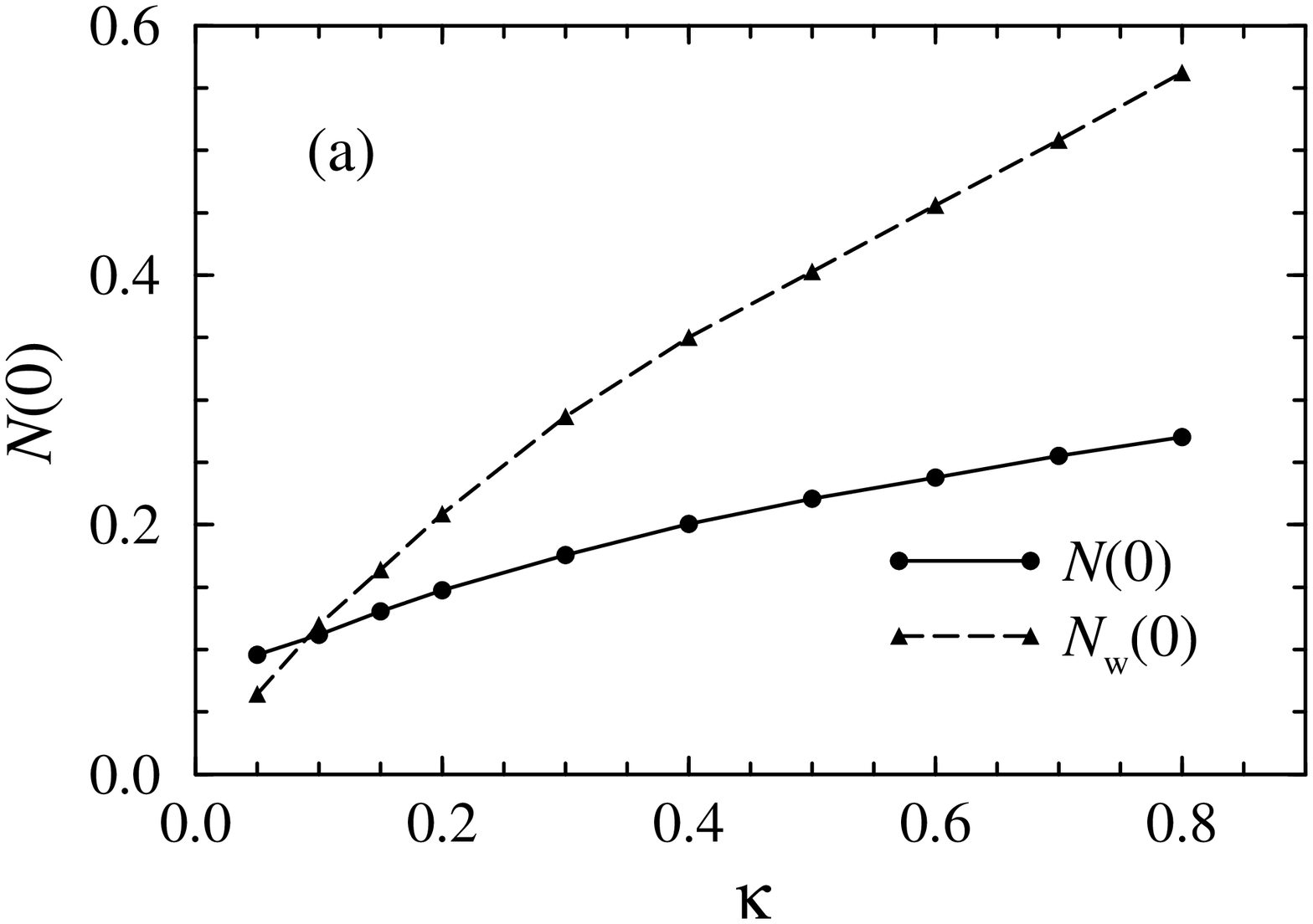}
\vskip -10truemm
\includegraphics[width=0.5\textwidth,angle=-90]{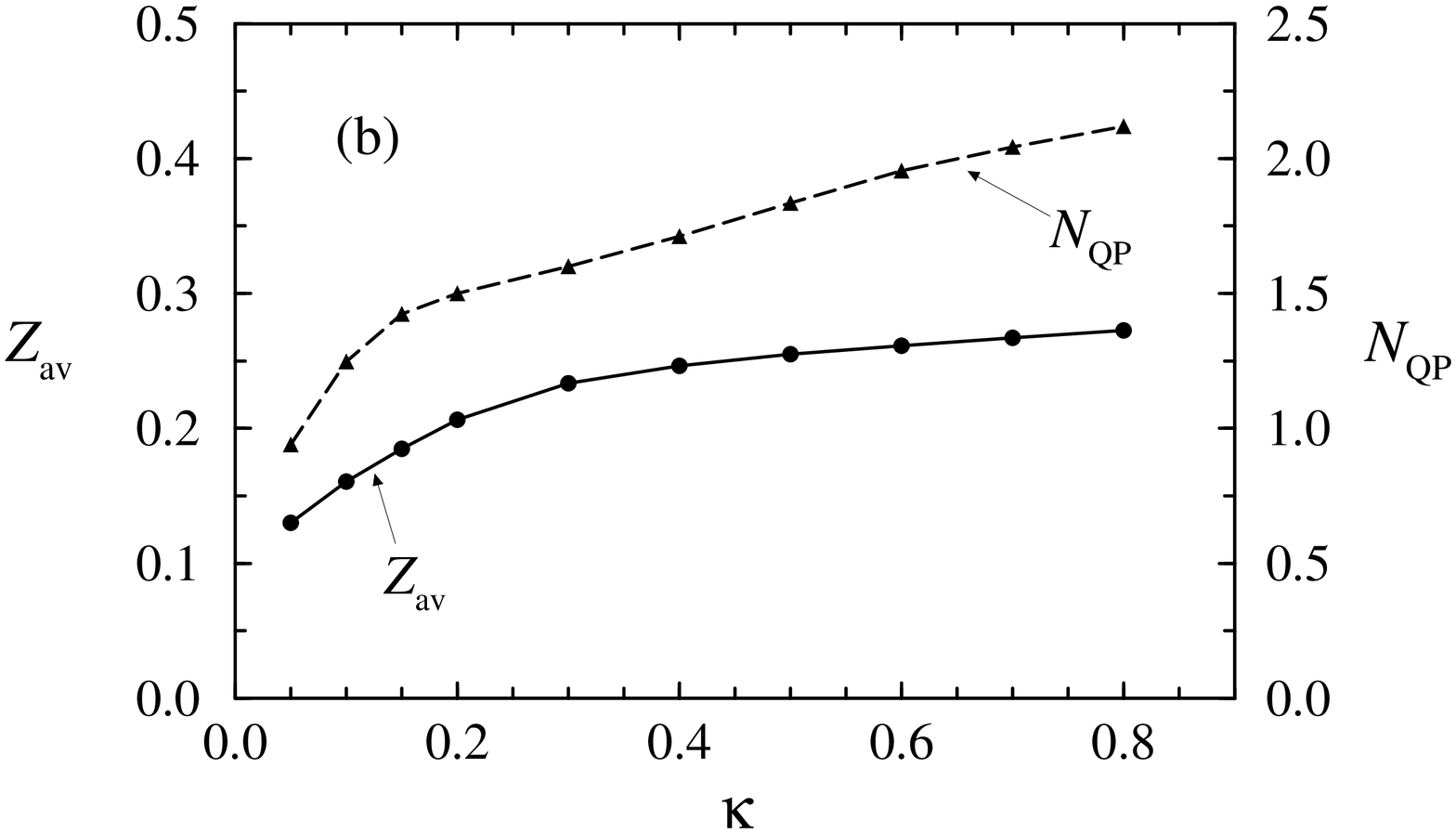}
\end{center}
\caption{(a) DOS ${\cal N}(0)$ and weighted DOS ${\cal
N}_{\rm w}(0)$ vs. $\kappa$.
(b) Average QP weight $Z_{\rm av}$ and QP DOS ${\cal N}_{\rm
QP}$ vs. $\kappa$. 
}
\label{fig6}
\end{figure}

It is also important to understand the role of finite $T>0$. The most
pronounced effect is on QP in the pseudogap part of the FS.  The main
conclusion is that weak QP peak with $Z_F \ll 1$ at $T=0$, as seen
e.g. in Fig.~5, is not just broadened but entirely disappears (becomes
incoherent) already at very small $T>T_s \ll J$, e.g. $T_s \sim
0.02~t$ for the situation in Fig.~5.  This can explain the puzzle that
ARPES experiments in fact do not observe any QP peak near $(\pi,0)$ in
the underdoped regime at $T>T_c$ \cite{mars,camp}.

\section{Conclusions}

In this paper we have presented our results for spectral functions and
the pseudogap within the $t$-$J$ model, which is the prototype model
for strongly correlated electrons and for superconducting cuprates in
particular. Here we first comment on the validity of the model and
results in a broader perspective of strange metals and materials close
to the metal-insulator transition. The physics of the $t$-$J$ model at
lower doping levels is determined by the interplay between the
magnetic exchange (dominating the undoped AFM insulator) and the
itinerant kinetic energy of fermions (being dominant at least in the
overdoped regime). Since itinerant fermions prefer a ferromagnetic
state, the quantum state at the intermediate doping is frustrated, the
quantum frustration showing up in large entropy, pronounced spin
fluctuations, non-Fermi liquid effects etc.  Evidently, this is one
path towards the metal-insulator transition, but definitely not the
only one possible. In this situation, fermionic and spin degrees of
freedom are coupled but both active and relevant for low-energy
properties. This is just the main content and assumption of the
presented theory for spectral functions and the pseudogap.

The EQM approach to dynamical properties seem to be promising since it
can treat exactly the constraint which is essential for the physics of
strongly correlated electrons. It has been recently also applied by
present authors to the analysis of spin fluctuations and collective
magnetic modes at low $T$. Our approximation for the self energy
within the EQM approach deals with the normal paramagnetic state and
treats the model as a coupled system (with derived effective coupling)
of fermions with spin fluctuations, where close to the AFM ordered
state both transverse and longitudinal spin fluctuations are
important.  Other contributions should be considered, e.g., the
coupling to pairing fluctuations, in order to treat the
superconducting state.

In this paper we present only the results of the simplified pseudogap
analysis. The results of full self-consistent treatment are
qualitatively similar \cite{prel1}. Within the present theory the
origin of the pseudogap feature is in the coupling to longitudinal
spin fluctuations near the AFM wavevector ${\bf Q}$ which determine
the QP properties in the 'hot' region, i.e. near the AFM zone
boundary.  The pseudogap opens predominantly in the same region and
its extent is dependent on $J$ and $t'$ but not directly on
$t$. Evidently the pseudogap bears a similarity to a d-wave-like
dependence along the FS (for $t'<0$) being largest near the $(\pi,0)$
point.  The strength of the pseudogap features depends mainly on
$\kappa$. It is important to note that apart from extremely small
$\kappa$ we are still dealing with a large FS.  Still, at $\kappa <
\kappa^* \sim 0.5$ parts of the FS near $(\pi/2,\pi/2)$ remain well
pronounced while the QP weight within the pseudogap part of the FS
are strongly suppressed, in particular near zone corners $(\pi,0)$.

The QP within the pseudogap have small weight $Z_F \ll 1$ but not
diminished (or even enhanced) $v_F$, which is the effect of the
nonlocal character of $\Sigma({\bf k},\omega)$. A consequence is that
QP within the pseudogap contribute much less to QP DOS ${\cal
N}_{QP}$. This could be plausible explanation of a well known
theoretical challenge that approaching the magnetic insulator both
DOS, i.e.  ${\cal N}(0)$ and ${\cal N}_{QP}$ vanish.

We presented results for $T=0$, however the extention to $T>0$ is
straightforward. Discussing only the effect on the pseudogap, we
notice that it is mainly affected by $\kappa$. So we can argue that the
pseudogap should be observable for $\kappa(c_h,T)<\kappa^* \sim
0.5$. This effectively determines the crossover temperature
$T^*(c_h)$. In the region of interest $\kappa$ is nearly linear in
both $T$ and $c_h$ so we would get approximately
\begin{equation}
T^* \sim T^*_0 (1- c_h/c_h^*), \label{eq24}
\end{equation}
where $T^*_0 \sim 0.6 J$ and $c_h^* \sim 0.15$. 

As described in previous sections, several features of our theory,
regarding the development of spectral functions, Fermi surface and
pseudogap, are at least qualitatively consistent with experimental results
of normal state properties in cuprates. However, further study within the
present formalism is necessary in order to explore possible
closer quantitative agreement with experiments as well as the 
emergence of superconductivity  within the same model.

\end{document}